\documentclass{ws-xxx}

\def\aprge{\buildrel > \over {_{\sim}}}
\newcommand{\ba}{\begin{eqnarray}}
\newcommand{\ea}{\end{eqnarray}}
\newcommand{\be}{\begin{equation}}
\newcommand{\ee}{\end{equation}}
\begin{document}
\title{Tracing Very High Energy Tau Neutrinos from Cosmological Sources in Ice}

\author{J. JONES, I. MOCIOIU, I. SARCEVIC} 

\address{Department of Physics \\
University of Arizona\\ 
Tucson AZ 85721\\
}

\author{M.~H.~Reno}
\address{
 Department of Physics and Astronomy\\ University of Iowa\\
Iowa City IA 52242}
\maketitle

\abstracts{
Astrophysical sources of ultrahigh energy neutrinos yield tau neutrino 
fluxes due to neutrino oscillations. 
We study in detail the contribution of tau neutrinos with energies 
above $10^6$ GeV relative to the contribution of the other flavors. 
We consider several different initial neutrino fluxes 
and include tau neutrino regeneration in transit through the Earth and 
energy loss of charged leptons. We discuss signals of tau neutrinos in 
detectors such as IceCube, RICE and ANITA. 
}

\section{Introduction}

The experimental evidence of $\nu_\mu\leftrightarrow\nu_\tau$ neutrino
oscillations \cite{2} means that astrophysical sources of muon neutrinos
become sources of $\nu_\mu$ and $\nu_\tau$ in equal proportions after
oscillations over astronomical distances \cite{3}. Although $\nu_\mu$ and
$\nu_\tau$ have identical interaction cross sections
at high energies, signals from
$\nu_\tau\rightarrow \tau$ conversions have the potential to contribute
differently from $\nu_\mu$ signals. The $\tau$ lepton can decay
far from the detector, regenerating $\nu_\tau$ \cite{4}. 
This also occurs with
$\mu$ decays, but electromagnetic energy loss coupled with the long
muon lifetime make the $\nu_\mu$ regeneration from muon decays irrelevant
for high energies. A second signal of $\nu_\tau\rightarrow\tau$ is
the tau decay itself \cite{5}. 

We have studied in detail the propagation of all flavors of 
neutrinos with very high energy ($E \geq 10^6$ GeV) as 
they traverse the Earth.  Because of the high energies attenuation shadows 
most of the upward-going solid angle at high energies, so we have limited
our consideration to nadir angles larger than $80^\circ$.
We are particularly interested in 
the contribution from tau neutrinos, 
produced in oscillations of extragalactic muon 
neutrinos as they travel large astrophysical distances.  

For most astrophysical sources, the neutrinos are produced in pion
decays, 
which determine the flavor ratio $\nu_e:\nu_\mu:\nu_\tau$ to be
$1:2:0$. 
After propagation over 
very long distances, neutrino oscillations change this ratio to
$1:1:1$ 
because of the 
maximal $\nu_\mu\leftrightarrow\nu_\tau$ mixing. 
For the GZK flux, $\nu_e$ and $\nu_\mu$ incident
fluxes are different because of the additional contributions from 
$\bar{\nu}_e$ from neutron decay and $\nu_e$ from $\mu^+$ decays
\cite{GZK}. 
Because of 
this, the flavor ratio at Earth is affected by the full three flavor
mixing 
and is 
different from $1:1:1$. Given fluxes at the source 
$F^0_{\nu_e}$, 
$F^0_{\nu_\mu}$ and $F^0_{\nu_\tau}$, the fluxes at Earth become:
\ba
F_{\nu_e}&=&F^0_{\nu_e}-\frac{1}{4}\sin^22\theta_{12} (2 F^0_{\nu_e}-F^0_{\nu_\mu}-
F^0_{\nu_\tau})
\label{fle}
\\
F_{\nu_\mu}&=&F_{\nu_\tau}=\frac{1}{2}(F^0_{\nu_\mu}+F^0_{\nu_\tau})+\frac{1}{8}\sin^22\theta_{12}
(2 F^0_{\nu_e}-F^0_{\nu_\mu}-F^0_{\nu_\tau})
\label{flmutau}
\ea
where $\theta_{12}$ is the mixing angle relevant for solar neutrino oscillations. We 
have assumed that $\theta_{23}$, the mixing angle relevant for atmospheric neutrino 
oscillations, is maximal and $\theta_{13}$ is very small, as shown by reactor experiments,
as well as atmospheric and solar data. 

Z burst neutrinos\cite{zburst} from models with ultrahigh energy
neutrinos scattering with relic neutrinos to produce Z bosons is another
neutrino flux considered below, where neutrino mixing yields flux ratios
of $1:1:1$.

\begin{figure}[t]
\centerline{\epsfxsize=3.5in\epsfbox{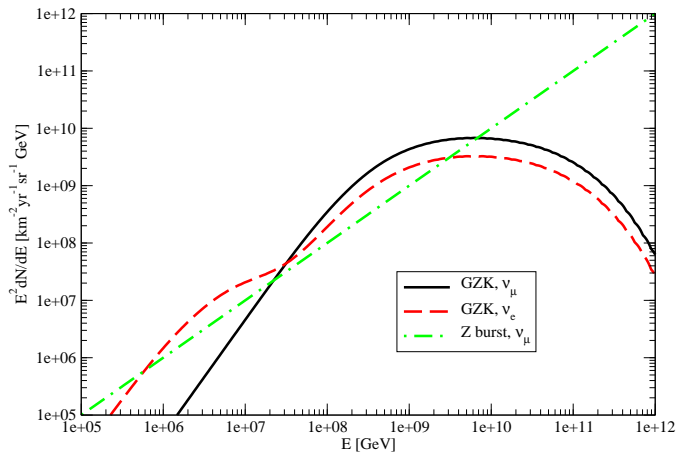}}   
\caption{Initial Neutrino Fluxes}
\label{fluxes}
\end{figure}

The initial fluxes for GZK and Z burst neutrinos are shown in Fig. \ref{fluxes}.

In our propagation of neutrinos and charged leptons through the Earth\cite{1}, 
we have focused on kilometer-sized neutrino detectors, such as 
ICECUBE~\cite{icecube} and the Radio Ice Cerenkov
Experiment (RICE)\cite{rice} and 
on a detector with much larger effective area which uses Antarctic 
ice as a converter, the Antarctic Impulsive Transient Antenna
(ANITA)\cite{anita}.

\section{Neutrino Propagation}

Attenuation and regeneration of neutrinos and charged leptons are
governed by the following transport equations:
\ba
\frac{\partial F_{\nu_{\tau}}(E,X)}{\partial X}\!\! 
&\!\!\!\!=\!\!\!\!&\!\!-
N_A\sigma^{tot}(E) {F_{\nu_{\tau}}(E,X)}
+ N_A\int_E^\infty dE_y F_{\nu_{\tau}}(E_y,X)\frac{d\sigma^{NC}}{\!\!\!\!\!dE}
 (E_y,E)
\nonumber \\
&& + \int_E^\infty dE_y \frac{F_{\tau}(E,X)}{\lambda_\tau^{dec}}
\frac{dn}{dE}(E_y,E)
\label{nuprop}
\ea
\be
 \frac{\partial F_\tau(E,X)}{\partial X}=  
        - \frac{F_\tau(E,X)}{\lambda_\tau^{dec}(E,X,\theta)}
+ N_A 
\int_E^\infty dE_y F_{\nu_{\tau}}(E_y,X)\frac{d\sigma^{CC}}{\!\!\!\!\!dE} 
(E_y,E)
\label{tauprop}
\ee
\be
-\frac{dE_\tau}{dX}=\alpha+\beta E_\tau
\label{eloss}
\ee
Here 
$F_{\nu_{\tau}}(E,X)=dN_{\nu_\tau}/dE$ and $
 F_\tau(E,X)=dN_\tau/dE$ are the differential energy
spectra of tau neutrinos and taus
respectively, for lepton energy $E$, 
at a column depth $X$ in the medium defined by
\be
X = \int_0^L\rho(L')dL'.
\ee

For tau neutrinos, we take into account the attenuation by charged current 
interactions, the shift in energy due to neutral current interactions and the
regeneration from tau decay. For tau leptons we consider their production in 
charged
current $\nu_\tau$ interactions, their decay, as well as 
electromagnetic energy loss.

The effective decay length of produced taus does not go above $10^7$ cm,
even for $E_\tau=10^{12}$ GeV.
This is because electromagnetic energy loss over that distance reduces the
tau energy to about $10^8$ GeV, at which point the tau is more likely 
to decay than interact electromagnetically \cite{10}.

We have found that the $\nu_\tau$ 
flux above $10^8$ GeV resembles the $\nu_\mu$ flux. 
The lore that the Earth is transparent to tau neutrinos is not applicable
in the high energy regime. Tau neutrino pileups at small angles with respect to
the horizon are significantly damped due to tau electromagnetic energy loss above
$E_\tau\sim 10^8$ GeV if the column depth is at least as large as the neutrino
interaction length.

At lower energies, $E \leq 
10^8$ GeV, regeneration of $\nu_\tau$ becomes important for trajectories 
where the other flavors of neutrinos are strongly
attenuated, but the $\nu_\tau$ regeneration is very effective. 
The regeneration effect depends strongly on the shape of the initial 
flux and it is larger for flatter fluxes.
The enhancement due to regeneration 
also depends on the amount of 
material traversed by neutrinos and leptons, i.e. on 
nadir 
angle. For GZK neutrinos, we have found that 
the enhancement peaks between $10^6$ and a few$\times 10^7$ GeV depending on
trajectory.

\begin{figure}[t]
\centerline{\epsfxsize=3.5in\epsfbox{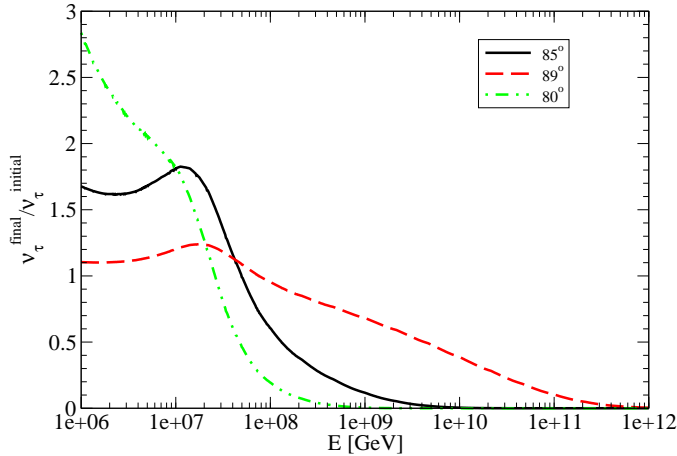}}   
\caption{Ratio $\nu_\tau/\nu_\mu$ for GZK neutrinos, 
at nadir angles of $85^\circ$ and $89^\circ$.}
\label{fig:ratio}
\end{figure}    
   
Fig. \ref{fig:ratio} shows the ratio of the tau neutrino flux after
propagation  to incident tau neutrino flux, for $89^\circ$,
$85^\circ$and 
$80^\circ$.
This ratio illustrates a combination of the 
regeneration of $\nu_\tau$ due to tau decay and the attenuation of all 
neutrino fluxes. 
For $89^\circ$, where both the total distance and the density are smaller, 
the attenuation is less dramatic, and the flux can be significant
even at high energy. The regeneration in this case can add about $25\%$ 
corrections at energies between $10^7$ and $10^8$ GeV. 
For $85^\circ$ the relative enhancement is around $80\%$ and peaked at 
slightly lower 
energies, while at $80^\circ$ it is almost a factor of 3 at low energy. At 
$80^\circ$, however, the flux is very strongly attenuated for energies above 
a few $\times 10^7$ GeV. 
It is already clear from here that the total rates will be dominated by the 
nearly horizontal trajectories 
that go through a small amount of matter. However, rates can get significant 
enhancements at low energies where the regeneration from tau decays adds an 
important contribution even for longer trajectories.

\section{Showers}

\begin{figure}[t]
\centerline{\epsfxsize=3.5in\epsfbox{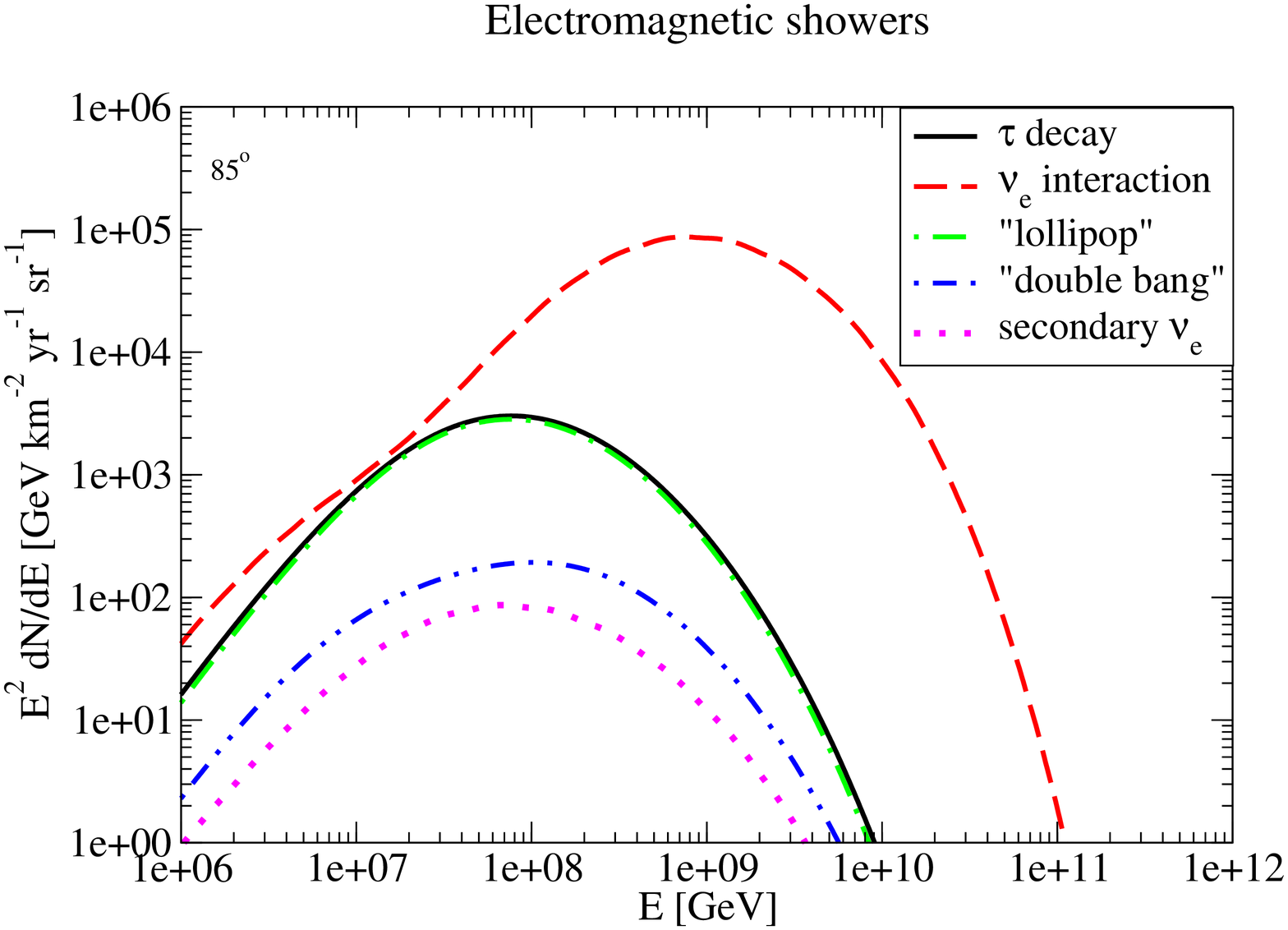}}   
\caption{Electromagnetic showers for GZK neutrinos \label{em}}
\end{figure}

\begin{figure}[t]
\centerline{\epsfxsize=3.5in\epsfbox{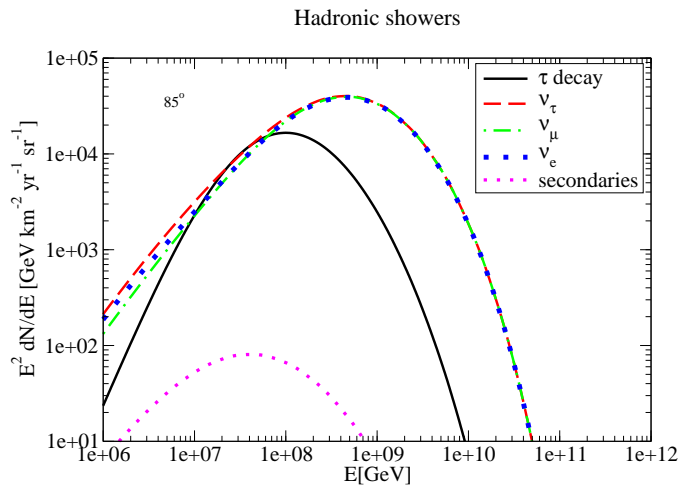}}   
\caption{Hadronic showers for GZK neutrinos \label{had}}
\end{figure}

We have translated the neutrino fluxes and tau lepton fluxes into rates for 
electromagnetic and hadronic showers at selected angles to see the effect
of attenuation, regeneration, and the different energy dependences of the
incident fluxes.  We have focused on comparing the $\nu_\tau$ contribution
to the $\nu_e$ and $\nu_\mu$ contributions to determine in what range, if any,
$\nu_\tau$'s enhance shower rates. Electromagnetic shower distributions 
for a nadir angle of $85^\circ$ are shown in Fig. \ref{em}, while 
Fig. \ref{had} shows 
hadronic showers.

\begin{figure}[t]
\centerline{\epsfxsize=3.5in\epsfbox{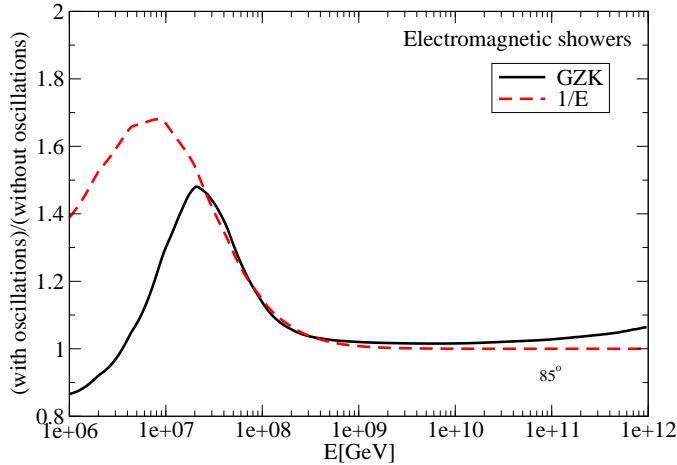}}  
\caption{Ratio of electromagnetic shower rates in the presence
and absence of 
$\nu_\mu\to\nu_\tau$ oscillations for GZK and $1/E$ neutrino spectra 
for nadir angle $85^\circ$ for a km size detector.}
\label{fig:shratioem}
\end{figure}

Fig. \ref{fig:shratioem} shows the ratio of the electromagnetic 
shower rates at nadir angle $85^\circ$ in the presence and absence of 
oscillations for the GZK and $Z$ burst 
neutrino fluxes (which have a characteristic
$1/E$ energy dependence). In absence of oscillations, the only contribution to 
electromagnetic showers comes from $\nu_e$ interactions. In the presence of 
$\nu_\mu\to\nu_\tau$ oscillations, electromagnetic decays of taus from tau 
neutrinos add significant contributions to these rates at energies below 
$10^8$ GeV. In the same time, for the GZK flux, $\nu_e\to\nu_{\mu,\tau}$ 
oscillations
reduce the number of $\nu_e$'s at low energy, such that below a few 
$\times 10^6$ GeV there are fewer electromagnetic showers than in the absence 
of oscillations.

The $\nu_\tau$ flux 
enhancements depend on the shape of the initial flux. The electromagnetic
showers are more sensitive to this shape than hadronic ones. The relative 
enhancement in hadronic showers is also smaller than for the electromagnetic 
showers. This is because for the electromagnetic signal the only contribution 
in the absence of taus is from electron neutrinos, while for hadrons the
tau contribution is compared to a much larger signal, from the interactions
of all flavors of neutrinos.
We have included contribution from secondary neutrinos, which 
we find to be relatively small for all fluxes.  

For kilometer-sized detectors, at for example a nadir angle of $85^\circ$, 
the maximal enhancement due to 
$\nu_\tau$ contribution to electromagnetic shower rates  
for the GZK flux is 
about $50\%$ at $3 \times 10^7$ GeV, while for a $1/E$ flux, it is even 
larger, 
about $70\%$, 
at slightly lower energy. These energy ranges are relevant for IceCube,
but not for RICE. For energies relevant to RICE,
tau neutrinos do not offer any appreciable
gain in electromagnetic shower signals compared to $\nu_e\rightarrow e$ CC
interactions, and they contribute at essentially the same level as $\nu_\mu$
to hadronic shower rates through NC interactions.

One of the reasons that tau neutrinos do not contribute large signals
to kilometer-sized detectors at very high energies is that high energy tau 
decay lengths
are very large, so the probability of a tau decaying in the detector is low.
For detectors like ANITA which can sample long trajectories through the
ice, one would expect a larger tau neutrino contribution to the signal from 
tau decay.
Despite the long trajectory (222 km with a maximum depth of 1 km for a
neutrino incident at $89^\circ$ nadir angle) the tau contributions to
the electromagnetic shower rate is quite small for fluxes expected to 
contribute
in the ANITA signal. 
For hadronic showers, the suppression of $\tau$ decay 
to hadrons
relative to $\nu_e$ NC interaction contributions is about the same
as for electromagnetic showers compared to $\nu_e\rightarrow e$. The $\nu_\tau$ contribution to the
hadronic shower rate from interactions is about the same as the $\nu_e$ contribution.
In summary, for ANITA, tau neutrinos do not give any additional signal beyond
what one would evaluate based on no regeneration from $\nu_\tau\rightarrow \tau
\rightarrow \nu_\tau$ due to tau electromagnetic energy loss at $E\aprge 10^8$ GeV.

\section*{Acknowledgments}

This work was supported in part by the Department of Energy under
contracts DE-FG02-91ER40664, DE-FG02-95ER40906 and DE-FG02-93ER40792. 


\end{document}